\documentclass[12pt]{article}

\usepackage{psfig}
\usepackage{latexsym}
\usepackage{amssymb}
\usepackage{epsfig,amsmath,graphics}

\newfont{\kreuz}{msbm10 scaled\magstep1}
\newfont{\Deutsch}{eufb10 scaled\magstep1}
\newfont{\deutsch}{eufb10}
\newfont{\schreib}{eusm10 scaled\magstep1}

\begin{document}

\newcommand{\be}{\begin{equation}}
\newcommand{\ee}{\end{equation}}
\newcommand{\ba}{\begin{array}}
\newcommand{\ea}{\end{array}}
\newcommand{\bea}{\begin{eqnarray}}
\newcommand{\eea}{\end{eqnarray}}
\newcommand{\bma}{\begin{matrix}}
\newcommand{\ema}{\end{matrix}}
\newcommand{\bpm}{\begin{pmatrix}}
\newcommand{\epm}{\end{pmatrix}}
\newcommand{\nn}{\nonumber}

\begin{titlepage}
\begin{center}



\vskip 5mm

{\large\bf $N=2$ Nonlinear Sigma Models in $N=1$ Superspace:\\
Four and Five Dimensions}
\vskip .4in

{\bf Jonathan Bagger and Chi Xiong} \\
\vskip .2in
{Department of Physics and Astronomy, Johns Hopkins University,} \\
{3400 North Charles Street, Baltimore, MD 21218, USA}\\
\vskip .2in
{bagger@jhu.edu}\\
{xiongc@pha.jhu.edu}
\end{center}

\vskip .5in

\abstract{We formulate four-dimensional $N=2$ supersymmetric nonlinear sigma models in $N=1$ superspace.  We show how to add superpotentials consistent with $N=2$ supersymmetry.  We lift our construction to higher-dimensional spacetime and write five-dimensional nonlinear sigma models in $N=1$ superspace.}


\end{titlepage}
\vfill




\newpage

\section{Introduction}

In four dimensions, $N=1$ supersymmetric nonlinear sigma models have a natural description in terms of K\"ahler geometry.  The complex coordinates of the K\"ahler manifold can be identified with the complex scalar fields contained in $N=1$ chiral superfelds.  The kinetic terms are the highest components of an arbitrary real function, the K\"ahler potential $K$.  The Yukawa terms are the $\theta^2$ components of an arbitrary analytic function, the superpotential $P$ \cite{GGRS, WessBagger}.

Four-dimensional $N=2$ supersymmetric models have a similar geometric description.  For $N=2$, the scalar fields are the coordinates of a hyper-K\"ahler manifold \cite{AGDF}.  One can construct such models in terms of component fields \cite{BaggerWitten, deWit}, in projective superspace \cite{Projective}, and in harmonic superspace \cite{Harmonic}, with an infinite number of auxiliary fields.

In this paper we will take a different approach, and describe $N=2$ supersymmetric nonlinear sigma models in terms of $N=1$ superfields.  (This approach was pioneered in refs.~\cite{GGRS, HuKLR, HiKLR}.)  The second supersymmetry restricts the K\"ahler potential and superpotential so that the K\"ahler manifold becomes hyper-K\"ahler.  The construction is easily lifted to five dimensions, where its intrinsically four-dimensional nature is especially well-suited to bulk-plus-brane constructions.  

The plan of this paper is as follows.  In section 2 we present some mathematical background and construct the general $N=2$ nonlinear sigma model in terms of $N=1$ superfields.  In section 3 we illustrate our results with three examples, including the Eguchi-Hanson model and a large class of models obtained via the so-called ``c-map"  \cite{deWit, cecotti}.  In section 4 we show that an $N=2$ superpotential can be included whenever the hyper-K\"alher manifold has a tri-holomorphic isometry \cite{BPSwall}.  In section 5 we lift our construction to five dimensions, and present the most general coupling of five-simensional hypermultiplets in terms of $N=1$ superfields.  Discussion and conclusions follow in section 6.


\section{Four Dimensions}

\subsection{Hyper-K\"ahler manifolds in complex coordinates}

In this paper we use $N=1$ chiral and anti-chiral superfields $ (\Phi^a, \bar{\Phi}^{\bar b}) $ to construct the matter couplings of $N=2$ supersymmetric theories.  We work in terms of complex coordinates because the lowest components of chiral superfields are complex scalars.  For the $N=1$ matter coupling, the complex coordinates parameterize a K\"ahler manifold.  In addition to a complex structure, such manifolds have a hermitian metric and a closed two-form $ \omega, $  called the K\"ahler form.  The metric is the second derivative of a real scalar function $ K( \Phi,\bar{\Phi}), $ called the K\"ahler potential. 

In general, $N=2$ matter couplings require the complex manifold to be not just K\"ahler, but hyper-K\"ahler.  A $4n$ (real) dimensional Riemannian manifold $ (M, g) $ is hyper-K\"ahler if it has three independent complex structures $ (I, J, K) $ such that the metric $ g $ is K\"ahler with repect to each of them.  Consequently, it has three K\"ahler forms, $ (\omega_1, \omega_2, \omega_3) = (g I, g J, g K) $.  In what follows we will single out one of the three complex structures, say $I$, and choose $2n$ complex coordinates with respect to it.  We then write the other two complex structures as $ \omega^{\pm} = \omega_2 \pm i \omega_3, $ a nondegenerate, closed and (anti)holomorphic two-form \cite{Besse}.  This shows that a hyper-K\"ahler manifold is a K\"ahler manifold with an additional holomorphic symplectic structure.

To describe the $N=2$ supersymmetric models in $N=1$ superspace, we will start with the most general supersymmetric action in $N=1$ superspace.  We then make an ansatz for the second supersymmetry and demand that the action be invariant under it.  This constrains the K\"ahler metric, the superpotential, and the second supersymmetry transformation laws.  We will see that the constraints are satisfied precisely when the K\"ahler manifold is hyper-K\"ahler.

\subsection{$N=2$ nonlinear sigma models in four dimensions}

We start with an ordinary $N=1$ sigma model in $N=1$ superspace,
\begin{equation} \label{eq:4d}
S_4=
\int\!d^4x d^4 \theta ~ K ( \Phi, \bar{\Phi})\ ,
\end{equation}
where $K$ is the K\"ahler potential of a K\"ahler manifold, and we defer the discussion of the superpotential until section 4.  The first supersymmetry is manifest.  We take the second supersymmetry to be given by the following ansatz,\footnote{Our ansatz (\ref{eq:2nd}) is different from that of ref.~\cite{HuKLR}.  In our case, the second supersymmetry closes off shell, while in \cite{HuKLR} the equations of motion are needed.  For that reason our construction is more readily generalized to higher-dimensional spacetime.}
\bea \label{eq:2nd}
\nonumber
\delta_{\eta} \Phi^{a} &=& \frac{1}{2} \bar{D^2} [ N^a (\theta \eta +
\bar{\theta} \bar{\eta} ) ] \\
\delta_{\eta} \bar{\Phi}^b &=& \frac{1}{2} {D^2} [ {\bar{N}}^{\bar{b}}
(\theta \eta + \bar{\theta} \bar{\eta} ) ],
\eea
where $ N^a = g^{a \bar{b}} \bar{N}_{\bar{b}},$ $\bar{N}_{\bar{b}} = \bar{N}_{\bar{b}} ( \Phi, \bar{\Phi} )$.

To proceed further, we need to solve for the functions $N^a$.  Define $ \Omega_{ab} \equiv \nabla_a N_b = \partial_{a} N_b - \Gamma^{c}_{ab} N_c,$  together with its complex conjugate, $ \bar \Omega_{\bar{a}\bar{b}}. $  We find that the action (\ref{eq:4d}) is invariant under the transformation (\ref{eq:2nd}) if the $ \Omega_{ab} $ satisfy
\bea
& 1. &  \Omega_{ab} = - \Omega_{ba} \nonumber \\
& 2. &  \Omega_{ab} , ~ \bar{\Omega}_{\bar{a}\bar{b}} ~~ \textrm{are covariantly 
constant} \nonumber \\
& 3. &  \Omega^{a}{}_{\bar{c}} ~ \Omega^{\bar{c}}{}_{b} = -
\delta^{a}{}_{b}, \nonumber
\eea
where $ \Omega^{a}{}_{\bar{b}} \equiv g^{a \bar{c}} \bar{\Omega}_{\bar{c}\bar{b}}$ and $\Omega^{\bar{a}}{}_{b} \equiv g^{c \bar{a}} \Omega_{cb} $. Moreover, with these conditions, the commutator $ [\delta_{\eta_2},  \delta_{\xi_2} ] \Phi $ closes {\it without} using the equations of motion.  The commutator of the first and the second supersymmetries  $[\delta_{\eta_1}, \delta_{\xi_2} ] $ also closes, with the help of the equations of motion. 

The conditions 1 -- 3 above imply that $ \Omega_{ab},$ $ \bar{\Omega}_{\bar{a}\bar{b}}$ are the holomorphic and anti-holomorphic two forms $\omega^\pm$ defined on the hyper-K\"ahler manifold, as discussed in the previous section.  They also imply that 
\be
\Omega_{ab} = \frac{1}{2} (\partial_a N_b - \partial_b N_a).
\ee
The three complex structures can be written in terms of $ \Omega^a{}_{\bar{b}},$ and $ \Omega^{\bar{a}}{}_{b} $ as follows \cite{deWit, HuKLR},
\be
J^{1}= \left(\bma 0 & -i \Omega^{a}{}_{\bar{b}} \\ i \Omega^{\bar{a}}{}_{b}
&
0
\ema \right) ~~~~
J^{2}= \left(\bma 0 &  \Omega^{a}{}_{\bar{b}} \\  \Omega^{\bar{a}}{}_{b}  &
0
\ema
\right) ~~~~
J^{3}= \left(\bma  -i \delta^{a}{}_{{b}}  & 0 \\  0 & i
\delta^{\bar{a}}{}_{\bar{b}}
\ema \right).
\ee
They satisfy
\be
J^{A} J^{B} = - I \delta^{AB} -
\varepsilon^{ABC} J^{C},
\ee
as required on a hyper-K\"ahler manifold \cite{AGDF,BaggerWitten}.

The functions $ N^a,$  $N_b $ define the second supersymmetry transformations, so we would like to find explicit expressions for them.  After a little manipulation, it is not hard to check that
\be \label{eq:NaNb}
N^a = - K_b \Omega^{ba},  ~~~~ N_a = K_{\bar{b}} \Omega^{\bar{b}}{}_{a}.
\ee
From these expressions, we see that $ N^a$ and $N_b $ transform as follows
\be
N^a  \rightarrow  N^a  + F_b \Omega^{ba}, ~~~~
N_a  \rightarrow  N_a  + \partial_a ( \bar{N^b} \bar{F}_{\bar{b}})
\ee
under K\"ahler transformations,
\be
K ( \Phi, \bar{\Phi} )  \rightarrow K ( \Phi, \bar{\Phi} ) + F(\Phi) + \bar{F}(\bar{\Phi}).
\ee
The functions $ N^a $ shift by a holomorphic function, so the second supersymmetry transformations (\ref{eq:2nd}) do not change.  The function $ N_a $ shifts by a total derivative, so $\Omega_{ab} = \frac{1}{2} (\partial_a N_b - \partial_b N_a) $ is left invariant, consistent with eq.\ (\ref{eq:NaNb}).

\section{Examples}

In this section we present three examples. (These examples were studied using different approaches in \cite{GGRS, deWit, HuKLR, cecotti}.)  We start with the simplest case: the free hypermultiplet.  We then study the Eguchi-Hanson model, which is the contangent bundle on complex project space $ CP(1)$.  We finish by examining a large class of hyper-K\"ahler manifolds obtained via the ``c-map."

\subsection{Free Hypermultiplet}

We write the coordinates $ (\Phi^{1}, \Phi^{2}) = (X, Y) $, in which case the K\"ahler potential is
\be
K ( X, Y,\bar{X},\bar{Y} ) = \bar{X} X + \bar{Y} Y .
\ee
The K\"ahler metric is simply $g_{a \bar{b}}= \left(\bma  1  & 0 \\  0 & 1  \ema \right), $ and all
connections vanish.  The form $ \Omega_{ab} = \varepsilon_{ab} = \left(\bma  0  & -1 \\  1 & 0  \ema \right) $.  Therefore $ N^a = g^{a \bar{b}} \varepsilon_{\bar{b} \bar{c}} \bar{\Phi}^c $, and the supersymmetry transformation law is
\bea
\nonumber
\delta_{\eta} X = - \frac{1}{2} \bar{D^2} [ \bar{Y} (\theta
\eta + \bar{\theta} \bar{\eta} ) ] &,&
\delta_{\eta} Y = + \frac{1}{2} \bar{D^2} [ \bar{X} (\theta
\eta + \bar{\theta} \bar{\eta} ) ] \\
\delta_{\eta} \bar{X} = - \frac{1}{2} {D^2} [ Y
(\theta \eta + \bar{\theta} \bar{\eta} ) ] &,&
\delta_{\eta} \bar{Y} = + \frac{1}{2} {D^2} [ X
(\theta \eta + \bar{\theta} \bar{\eta} ) ].
\eea

\subsection{Eguchi-Hanson $ T^{*} CP(1)$ model}

The Eguchi-Hanson $ T^{*} CP(1)$ Model is one representative of a large class of hyper-K\"ahler manifolds, $ T^{*} CP(n)$.  The K\"ahler potential is
\cite{HuKLR}
\be
K (X, Y,\bar{X}, \bar{Y}  ) = \sqrt{1+ \rho^4} - Log \frac{1+\sqrt{1+ \rho^4}}{\rho^2}, ~~~~
\rho=\sqrt{X \bar{X} + Y \bar{Y}},
\ee
so the K\"ahler metric and its inverse are
\be
g_{a \bar{b}}=\frac{1}{\rho^4 \sqrt{1+\rho^4} }
\left(\bma  \rho^6 + Y \bar{Y}   & - Y \bar{X} \\
- X \bar{Y}  & \rho^6 + X \bar{X} \ema \right),
\ee
\be
g^{a \bar{b}}=\frac{1}{\rho^4 \sqrt{1+\rho^4} }
\left(\bma  \rho^6 + X \bar{X}   &  X \bar{Y} \\
Y \bar{X}  & \rho^6 + Y \bar{Y} \ema \right).
\ee
The metric and its inverse satisfy $ \Omega_{ab} g^{b \bar{c}} \bar{\Omega}_{\bar{c} \bar{d}} = - g_{a \bar{d}},$ with
\be
\Omega_{ab} = \bar{\Omega}_{\bar{a} \bar{b}} =
\left(\bma  0  & -1 \\  1 & 0  \ema \right), ~~~~~
\Omega^{ab} = \bar{\Omega}^{\bar{a} \bar{b}} =
\left(\bma  0  & 1 \\  -1 & 0  \ema \right).
\ee
The second supersymmetry transformation is defined by $ N^X$ and $N^Y$,
\bea
N^X = + \frac{\sqrt{1+ \rho^4}}{\rho^2} \bar{Y}, &  &
N^Y = - \frac{\sqrt{1+ \rho^4}}{\rho^2} \bar{X}.
\eea
One can check all the conditions given in section 2.2 are satisfied.

\subsection{C-map}

The ``c-map'' allows one to construct a hyper-K\"ahler manifold from the K\"ahler manifold associated with $ N=2 $ vector multiplets.  The coordinates are taken as $ (\Phi^1, \Phi^2, ... \Phi^{2n})=(X^1, X^2, ...X^n, Y_1, Y_2, ... Y_n) $. The K\"ahler potential is as follows  \cite{deWit, cecotti},
\be
K (X^I, Y_J, \bar{X}^I, \bar{Y}_J ) = G(X^I, \bar{X}^I) + \frac{1}{2} G^{I
\bar{J}} (X^I, \bar{X}^I) (Y + \bar{Y} )_I  (Y + \bar{Y} )_J,
\ee
where $G(X^I, \bar{X}^I)= X^I \bar{F}_{\bar{I}} (\bar{X}) + \bar{X}^{\bar{I}} F_I (X),$ $F_I (X)= \partial_I F,$ and $\bar{F}_{\bar{I}} =\partial_{\bar{I}} \bar{F}$. In these expressions, $ G^{I \bar{J}} $ is the inverse of $ G_{I \bar{J}}= F_{IJ} (X)+\bar{F}_{\bar{I} \bar{J}} $.

With these definitions, the K\"ahler metric can be written in matrix form,
\bea
\nonumber
g_{a \bar{b}} &=& \left(\bma  (G+ S G^{-1} \bar{S} )_{A \bar{B}} &
(-S G^{-1})_{A}^{~~\bar{B}}  \\  (- G^{-1} \bar{S})^{A}_{~~\bar{B}} &
(G^{-1})^{A \bar{B}}   \ema \right) \\[2mm]
&=&  \left(\bma    I & -S  \\  0 & I  \ema \right) ~
\left(\bma   I & 0  \\  0 & I  \ema \right) ~
\left(\bma   G & 0  \\  -G^{-1} \bar{S} & G^{-1}  \ema \right)_{a\bar{b}},
\eea
where $S_{IJ}=F_{IJK} G^{K\bar{L}} {(Y+\bar{Y})}_L$.  From the matrix expressions, we see that $ \det(g_{a \bar{b}})=1$ and that its inverse is
\be
g^{a \bar{b}}= \left(\bma  (G^{-1} )^{A \bar{B}} &
( G^{-1} \bar{S})^{A}_{~~\bar{B}}  \\  ( S G^{-1})_{A}^{~~\bar{B}} &
(G + S G^{-1} \bar{S})_{A \bar{B}}   \ema \right).
\ee
Comparing the metric and its inverse, we find $ \Omega_{ab} g^{b \bar{c}} \bar{\Omega}_{\bar{c} \bar{d}} = - g_{a \bar{d}} $,
with
\be
\Omega_{ab} = \bar{\Omega}_{\bar{a} \bar{b}} =\left(\bma  0  & I \\  -I & 0  \ema \right), 
\ee
The second supersymmetry transformation is specified by
\bea
N^a = \left\{ \begin{array}{ll}
 G^{A \bar{B}} (Y + \bar{Y})_{B}, & a=1,...n, \\
 -G_{A} + \frac{1}{2} S_{AB} G^{C\bar{B}} (Y + \bar{Y} )_C, &a=n+1,...2n.
\end{array} \right.
\eea

\section{Superpotentials}

In this section we show how to add a superpotential to the action while preserving $N=2$ supersymmetry.  The second supersymmetry places extra restrictions on the allowed superpotentials.  We start by writing the action as follows,
\begin{equation} \label{eq:4DP}
S_4=
\int\!d^4x d^4 \theta ~ K ( \Phi, \bar{\Phi} ) + \int d^2 \theta~
s\,P(\Phi)
+ \int d^2 \bar{\theta} ~s^{\ast}\,\bar{P}( \bar{\Phi}),
\end{equation}
where $s$ is a phase, $s = e^{i\sigma}$.  We change the second supersymmetry transformation law to\footnote{There are other possibilities.  For example, one may replace (\ref{eq:2ndp}) by $\delta_{\eta} \Phi^{a} =\frac{1}{2} \bar{D^2} [ N^a (\bar{\theta} \bar{\eta} ) ] - 2 s^{\ast} \Omega^{ab} P_b \theta \eta - 2 \Omega^{ab} [ -\frac{1}{4} \bar{D^2} K_b + s P_b ]\theta \eta $.  The third term  is proportional to the anti-symmetric tensor $ \Omega^{ab} $ times the equation of motion, so it can be dropped.  This leaves $ \delta_{\eta} \Phi^{a} =\frac{1}{2} \bar{D^2} [ N^a (\bar{\theta} \bar{\eta} ) ] - 2 s^{\ast}\Omega^{ab} P_b \theta \eta $.  The new transformation requires the tri-holomorphic conditions {\it and} the equations of motion to close the second supersymmetry, $ [\delta_{\eta_2}, \delta_{\xi_2} ] $.  The advantage of (\ref{eq:2ndp}) is that the closure of $ [\delta_{\eta_2},  \delta_{\xi_2} ] $ leads directly to the tri-holomorphic condition, without using the equations of motion. } 
\bea \label{eq:2ndp}
\nonumber
\delta_{\eta} \Phi^{a} &=& \frac{1}{2} \bar{D^2} [ N^a (\theta \eta +
\bar{\theta} \bar{\eta} ) ] - 2(s+s^{\ast}) \Omega^{ab} P_b \theta \eta \\
\delta_{\eta} \bar{\Phi}^a &=& \frac{1}{2} {D^2} [ {\bar{N}}^{\bar{a}}
(\theta \eta + \bar{\theta} \bar{\eta} ) ] - 2(s+s^{\ast}) \bar{\Omega}^{\bar{a} \bar{b}}
\bar{P}_{\bar{b}} \bar{\theta} \bar{\eta},
\eea
where $ N^a = -K_b \Omega^{ba} $.  The action is invariant provided
\be \label{eq:KP}
K_a  \Omega^{ab}  P_b- K_{\bar{a}} \bar{\Omega}^{\bar{a} \bar{b}}
 \bar{P}_{\bar{b}} = f(\phi) + \bar{f}(\bar{\phi}),
\ee
where $f(\phi)$ and $\bar{f}(\bar{\phi})$ are (anti)holomorphic functions.  Closure of the supersymmetry algebra on $ [\delta_{\eta_2},  \delta_{\xi_2} ] \Phi $ requires that
\be
\partial_{\bar a} (N^a{}_{,\bar{b}}\, \bar{\Omega}^{\bar{b} \bar{c}} \bar{P}_{\bar{c}} +
(\Omega^{ab}  P_b)_{,c}\, N^c - N^a{}_{,b}\, \Omega^{bc}  P_c) = 0,
\ee
without using the equations of motion.

Let us define a vector $ X^a \equiv i ~ \Omega^{ab} P_b $ that is manifestly holomorphic.  In terms of this vector, the above conditions become
\bea  \label{eq:killing}
& 1. &  \nabla_{a} \bar{X}_{\bar{b}} + \nabla_{\bar{b}} X_a =0 \nonumber \\  \label{eq:triholo}
& 2. &  \Omega^{a}{}_{\bar{c}}\, \nabla_{\bar{b}} \bar{X}^{\bar{c}}
- \Omega^{c}{}_{\bar{b}}\, \nabla_{c} X^a =0.
\eea
The first condition is just the Killing equation; it is obtained by taking the derivative of both sides of (\ref{eq:KP}).  It restricts $ X^a $ to be a Killing vector on the hyper-K\"ahler manifold.  The second condition implies that the vector must 
be tri-holomorphic \cite{AGDF, Projective, HuKLR, BPSwall}.  This means that the isometry generated by $X^a $ must leave the three complex structures invariant, i.e., the Lie derivative of the complex structures $ \mathcal{L}_{X+\bar{X}} J^A = 0 $.

The superpotential terms give rise to a scalar potential in the component action,
\be
V= g^{a \bar b} P_a \bar{P}_{\bar{b}} = -\Omega_{ca}  g^{a \bar b}
\bar{\Omega}_{\bar{b} \bar{d}} \bar{X}^{\bar{d}} X^c = g_{c \bar{d}}
\bar{X}^{\bar{d}} X^c.
\ee
We see that the scalar potential is just the norm of the tri-holomorphic Killing vector. This coincides with the conclusion in \cite{BPSwall}, obtained by a component field construction.

A straightforward calculation shows that the commutators of the first and second supersymmetry transformations on $ A^a $ and $ \psi^a $ are as follows,
\bea
[\delta_{\eta_2},  \delta_{\zeta_1} ] A^a &=& 2 i X^a ( s \bar{\zeta}_1  \bar
{\eta}_2 - s^{\ast}\zeta_1 \eta_2) \nonumber \\[1mm] 
[\delta_{\eta_2},  \delta_{\zeta_1} ] \psi^a_{\alpha} &=& 2 i \partial_{b}
X^a ~ \psi^b_{\alpha} (s \bar{\zeta}_1  \bar{\eta}_2 - s^{\ast}\zeta_1 \eta_2).
\eea
Note that $ \partial_{b} X^a \psi^b_{\alpha} $ is the $ \theta $ component of the superfield $ X^a (\Phi ) = X^a (A) + \sqrt{2} \theta \psi^b \partial_{b} X^a + \cdots $.  The two equations are just the lowest and the $\theta$ component of superfield equation,
\be \label{eq:Xch}
[\delta_{\eta_2},  \delta_{\zeta_1} ] \Phi^a = 2 i X^a (\Phi) (s \bar{\zeta}_1  \bar
{\eta}_2 - s^{\ast} \zeta_1 \eta_2) .
\ee
The two supersymmetries close into the triholomorphic Killing vector $X^a$.

If there is more than one such Killing vector, (i.e.\ the tri-holomorphic isometry group is non-Abelian, with $n$ tri-holomorphic Killing vectors $ X^a_A,$ $ A=1,2, ...n $), calculations show that one can simply replace $ s P \rightarrow 
 s^A P_A $ in the action (\ref{eq:4DP}) and the transformation laws (\ref{eq:2ndp}), provided $ s^A = s n^A $, where $ n^A $ is a real unit vector.  This condition follows the fact that the variation $ s^A P_{Aa} (s^B + s^{*B}) P_{Bb} \Omega^{ab} $ must vanish; it leads to the condition $ s^A s^{*B} = s^B s^{*A} $, which is solved by $ s^A = s n^A $.  In other words, for general tri-holomorphic isometries, one simply replaces $ s P $ by $ s n^A P_A $.
 
As an example, let us consider the Eguchi-Hanson model.  Its  metric, complex structures and second supersymmetry transformations were given in section 3.2. To distinguish coordinates from Killing vector symbols, we denote the coordinates by $ (x, y) $. It is not hard to find the holomorphic Killing vectors in the $ (\partial_x, \partial_y ) $ basis:
\be
X_0= m_0 (ix, i y), ~~~~X_1= m_1 (ix, -i y), ~~~~ X_2 = m_2 (iy, ix ), ~~~~ X_3 = m_3 (y, -x),
\ee
where the $m_i$ are included for dimensional reasons.
$X_0 $ generates a U(1) group while $ X_1, X_2 $ and $ X_3 $ generate an SU(2). However, $ X_0 $ does not satisfy the tri-holomorphic condition (\ref{eq:triholo}), while the other three do. Therefore only $ X_1, X_2 $ and $ X_3 $ are allowed to be superpotentials.  The corresponding superpotentials are \cite{HuKLR}
\be
P_1= m_1\,xy , ~~~~ P_2 = \frac{m_2}{2} ( y^2-x^2), ~~~~ P_3 = - \frac{im_3}{2} (y^2+x^2).
\ee
The three superpotentials can be added to the action separately, or as a linear combination, as described above.  (The $ (X_1, P_1) $ case was also studied in \cite{ANNS} by introducing an auxiliary vector multiplet.)

\section{Five Dimensions}

Constructions with four-dimensional $N=1$ superfields are especially well-suited to five-dimensional bulk-plus-brane scenarios in which branes foliate the fifth dimension.  It is natural, then, to realize the five-dimensional bulk supersymmetry in terms of four-dimensional superfields.  Manifest Lorentz invariance is lost, but computational simplicity often more than makes up the difference.

From this point of view, each four-dimensional superfield is now a function of $x^5$, the coordinate in the fifth dimension.  As far as the two four-dimensional supersymmetries are concerned, the extra coordinate is just a label \cite{AGW}.  Each of the supersymmetries closes into a four-dimensional translation.  The full five-dimensional supersymmetry manifests itself in the closure of the first and second supersymmetries:  by Lorentz invariance, they must close into a translation along the fifth dimension.

To see how this works, let us consider the five-dimensional action,
\begin{equation} \label{eq:5d}
S_5=
\int\!d^5x d^4 \theta ~ K ( \Phi, \bar{\Phi} ) + \int d^2 \theta~
P(\Phi, \partial_5 \Phi )   ~+~ h.c.
\end{equation}
We take the second supersymmetry transformations from (\ref{eq:2nd}), and calculate the commutator of the first and second supersymmetry transformations on the field $A^a$.  We find
\be
[\delta_{\eta_2},  \delta_{\zeta_1} ] A^a = -2 \Omega^a_{~\bar{c}} ~
( F^{c*} - \frac{1}{2} \Gamma^{\bar{c}}_{\bar{b} \bar{d}} \bar{\psi}^d
\bar{\psi}^b ) ( \bar{\zeta}_1  \bar{\eta}_2 + \zeta_1 \eta_2).
\ee
By Lorentz invariance, we need to require
\be \label{eq:5scalar}
[\delta_{\eta_2},  \delta_{\zeta_1} ] A^a = 2 \partial_5 A^a ( \bar{\zeta}_1  \bar{\eta}_2 + \zeta_1 \eta_2),
\ee
which implies
\be \label{eq:aux}
g_{a\bar{c}}( F^{c*} - \frac{1}{2} \Gamma^{\bar{c}}_{\bar{b} \bar{d}} \bar{\psi}^d
\bar{\psi}^b ) =   \Omega_{ab} \partial_5 A^b.
\ee

Equation (\ref{eq:aux}) is the lowest component of the following superfield equation,
\be
- \frac{1}{4} \bar{D^2} K_a = \Omega_{ab} \partial_5 \Phi^b.
\ee
The five-dimensional superfield equation of motion is
\be
 \frac{1}{4} \bar{D^2} K_a = \frac{\partial P }{\partial \Phi^a} -
\frac{\partial}{\partial_5} \frac{\partial P }{\partial \partial_5 \Phi^a}.
\ee
Combining these results, we see that the supersymmetry algebra is Lorentz-invariant, provided
\be \label{eq:5dcons}
\frac{\partial P }{\partial \Phi^a} -
\frac{\partial}{\partial_5} \frac{\partial P }{\partial \partial_5 \Phi^a}
= - \Omega_{ab} \partial_5 \Phi^b.
\ee
This is a constraint on $P$.  With this constraint, the action (\ref{eq:5d}) is invariant under the full Lorentz symmetry, as well as under the transformation laws (\ref{eq:2nd}).

We have also checked the closure of algebra on the fermionic fields $\psi^a$.  Solving (\ref{eq:aux}) for $ F^a $ and plugging it into the component form of (\ref{eq:2nd}), we obtain
\be \label{eq:5spinor}
[\delta_{\eta_2},  \delta_{\zeta_1} ] \psi^a = 2 \partial_5 \psi^a 
( \bar{\zeta}_1  \bar{\eta}_2 + \zeta_1 \eta_2).
\ee
Equations (\ref{eq:5scalar}) and (\ref{eq:5spinor}) are the lowest and the $\theta$ components of superfield equation 
\be
[\delta_{\eta_2},  \delta_{\zeta_1} ] \Phi^a = 2 \partial_5 \Phi^a 
( \bar{\zeta}_1  \bar{\eta}_2 + \zeta_1 \eta_2),
\ee
respectively.

We are now ready to solve the constraint (\ref{eq:5dcons}) and find explicit expressions for the superpotential.  In the three previous examples, $ \Omega_{ab} $ was a constant matrix, so it is easy to confirm that $ P =\frac{1}{2} \Omega_{ab} \Phi^a \partial_5 \Phi^b $ gives a solution to (\ref{eq:5dcons}).  In the general case, we find that $ P = H_a ( \Phi ) \partial_5 \Phi^a $ solves the constraint (\ref{eq:5dcons}), so long as $ \Omega_{ab} = H_{a,b} - H_{b,a} $.

From the previous section, we know that $\Omega_{ab} = \frac{1}{2} (N_{b,a} - N_{a,b})$.  Therefore we may expect that the holomorphic functions $ H_a ( \Phi ) $ are related to the non-holomorphic functions $ N_a( \Phi, \bar{\Phi})$.  In fact, one may obtain an expression for $ H_a ( \Phi ) $ by integrating over the anti-holomorphic fields in $ N_a ( \Phi, \bar{\Phi} ) $, i.e.
\be \label{eq:hol}
 H_a ( \Phi ) \equiv -\frac{\int_{V} N_a ( \Phi, \bar{\Phi} )
 ~L(\bar{\Phi}) d \bar{\Phi} } 
 { 2 \int_{V} L(\bar{\Phi}) d \bar{\Phi} },
\ee
where $ L(\bar{\Phi}) $ is an integration kernel that makes the integral converge and $ d \bar{\Phi} \equiv d \bar{\Phi}^1 d \bar{\Phi}^2  \cdots d \bar{\Phi}^{2n} $.  It is easy to check that $ H_{a,b} - H_{b,a} = \Omega_{ab} $ since $ \Omega_{ab} $ is holomorphic,
\bea
H_{a,b} - H_{b,a}&=& -\frac{\int_{V} (N_{a,b}-N_{b,a})
 ~L(\bar{\Phi}) d \bar{\Phi} } 
 { 2 \int_{V} L(\bar{\Phi}) d \bar{\Phi}} \\ \nonumber
                 &=& -\frac{\int_{V} -2 \Omega_{ab}( \Phi)
 ~L(\bar{\Phi}) d \bar{\Phi}} 
 { 2 \int_{V} L(\bar{\Phi}) d \bar{\Phi}} \\ \nonumber
                  &=& \Omega_{ab}.
\eea 
Equation (\ref{eq:hol}) gives a solution to (\ref{eq:5dcons}).  Different solutions can be found by choosing different integration kernels $ L(\bar{\Phi}) $.  The component action, however, remains the same.

An alternate and simpler way is to replace the non-holomorphic fields by constants, 
\be \label{eq:hol2}
 H_a ( \Phi ) \equiv -\frac{1}{2} N_a ( \Phi, \bar{\Phi} ) {\Big |}_{\bar{\Phi}=c},
\ee
so $ H_{a,b} - H_{b,a} = -\frac{1}{2} (N_{a,b} ( \Phi, \bar{\Phi} ) {\big |}_{\bar{\Phi}=c}- N_{b,a} ( \Phi, \bar{\Phi} ) {\big |}_{\bar{\Phi}=c} ) = \Omega_{ab} $.   Equations (\ref{eq:hol}) and (\ref{eq:hol2}) can also be considered as proofs of the existence of $  H_a ( \Phi ) $.  For any hyper-K\"ahler manifold, once the complex structures and complex sympletic structures are found, we can calculate $ N_a = K_{\bar{b}} \Omega^{\bar{b}}_{~a} $ and then calculate $ H_a $ via (\ref{eq:hol}) or (\ref{eq:hol2}).

In order to demonstrate the five-dimensional Lorentz invariance of the action (\ref{eq:5d}), we write it in terms of component fields and integrate out the auxiliary fields through their equations of motion.  We define our five-dimensional spinors as follows,
\be \label{eq:5dspinor}
\Psi^a \equiv \left( \psi^a,  \Omega^a_{~\bar{c}} \bar{\psi}^c \right)^T, ~~~ 
\bar{\Psi}^{b} \equiv ( \Omega^{\bar{b}}_{~d} \psi^d, \bar{\psi}^b ),~~~ a,b = 1, 2, ...2n,
\ee
where the five-dimensional $ \Gamma $-matrices are
\be
\Gamma^M =
\left( \left(\bma  0  & \sigma^m \\  \overline{\sigma}^m & 0  \ema \right), ~~ 
  \left(\bma  -i  & 0 \\  0  & i  \ema \right) \right), ~~~\{\Gamma^M, \Gamma^N\} = -2 \eta^{MN}, 
  ~~\eta^{MN} = \{-1,1,1,1,1\}.
\ee
With these conventions, the action (\ref{eq:5d}) is fully Lorentz invariant,
\bea
\nonumber
S_5^{\mbox{\tiny{A}}} &=& - g_{a\bar{b}} \partial_M A^a \partial^M A^{b*}  
-i g_{a\bar{b}} \bar{\Psi}^{b} \Gamma^M (\partial_M \Psi^a + \Gamma^a_{cd} 
\partial_M A^d \Psi^c ) \\[2mm]
&& - \frac{1}{40} R_{a \bar{b} c \bar{d}} 
(\bar{\Psi}^{b} \Gamma_M  \Psi^a )(\bar{\Psi}^{d} \Gamma^M  \Psi^c ).
\eea
Note that the $ \Omega $ tensors do not appear explicitly in the action.  

It is also possible to include a superpotential in the five-dimensional theory.  The easiest way is to define $ P_{\textrm{new}} \equiv H_a (\Phi) \partial_5 \Phi^a + s P(\Phi) $.  It is not hard to show that the new five-dimensional action, with the superpotential, is invariant under the second supersymmetry transformation (\ref{eq:2ndp}), provided $X^a=i \Omega^{ab} P_b $ is a tri-holomorphic Killing vector, as in previous section.  Note that when we write the five-dimensional action in component field form, we find that the superpotential ``mass" terms are Lorentz-invariant if and only if the phase $s=\pm1$: 
\be
  -\frac{s}{2}\, \nabla_b\, \partial_a P ~\psi^b \psi^a -\frac{s^*}{2}\, \nabla_{\bar{b}}\, \partial_{\bar{a}} \bar{P} 
  ~\bar{\psi}^b \bar{\psi}^a = \pm \frac{i}{4} (\nabla_{\bar{a}} X_b - \nabla_b \bar{X}_{\bar{a}})~\bar{\Psi}^a \Psi^b,
\ee
when $s=\pm1$. Likewise, closure of the five-dimensional supersymmetry algebra also requires $s=\pm1$.  The component results might also be obtained through a nontrivial dimensional reduction of a six-dimensional massless hyper-K\"ahler multiplet, as discussed in \cite{BPSwall}. 

\section{Conclusions}

In this paper we presented a general construction of the four-dimensional $N=2$ nonlinear sigma model in $N=1$ superspace.  The second supersymmetry transformation laws were not unique; our ansatz made the second supersymmetry close without using the equations of motion.  The first and the second supersymmetries, however, did not close off-shell; they still required the equations of motion.

With our construction, it was straightforward to include central charges in the supersymmetry algebra. We chose different types of central charges -- either including a superpotential in four dimensions, or lifting the whole construction to five dimensions.   The five-dimensional setup is well-suited to model building in the brane-world scenario.

After this paper was submitted to the archive, the results of section 4 were obtained in projective superspace \cite{Kuzenko}.  Likewise, the six-dimensional supersymmetric nonlinear sigma model, formulated in terms of four-dimensional $N=1$ superspace, was presented in \cite{6D}.

This work was supported by the National Science Foundation, grant NSF-PHY-0401513.

\end{document}